\documentclass[draft]{agujournal2019}
\usepackage{url} 
\usepackage{lineno}
\usepackage[inline]{trackchanges} 
\usepackage{soul}
\usepackage{amsmath,amssymb,amsfonts}%

\draftfalse

\journalname{Journal of Advances in Modeling Earth Systems}

\begin{document}

%
%


\title{CERA: A Framework for Improved Generalization of Machine Learning Models to Changed Climates}

%
%




\authors{Shuchang Liu\affil{1}, Paul A. O'Gorman\affil{1}}

\affiliation{1}{Department of Earth, Atmospheric, and Planetary Sciences, Massachusetts Institute of Technology}




\correspondingauthor{Shuchang Liu}{shuchang.liu@hotmail.com}



\begin{keypoints}
\item CERA improves generalization across climates via climate-invariant representations, outperforming raw-input and physics-informed baselines.
\item Latent alignment using Earth Mover’s Distance improves both predictive accuracy and robustness across random seeds.
\item The framework is tested on a parameterization of moist processes, but has potential for other climate applications such as downscaling.
\end{keypoints}

%
%

%
%


\begin{abstract}
Robust generalization under climate change remains a major challenge for machine learning applications in climate science. Most existing approaches struggle to extrapolate beyond the climate they were trained on, leading to a strong dependence on training data from model simulations of warm climates. Use of climate-invariant inputs improves generalization but requires challenging manual feature engineering. Here, we present CERA (Climate-invariant Encoding through Representation Alignment), a machine learning framework consisting of an autoencoder with explicit latent-space alignment, followed by a predictor for downstream process estimation. We test CERA on the problem of parameterizing moist-physics processes. Without training on labeled data from a $+4\,\mathrm{K}$ climate, CERA leverages labeled control-climate data and unlabeled warmer-climate inputs to improve generalization to the warmer climate, outperforming both raw-input and physically informed baselines in predicting key moisture and energy tendencies. It captures not only the vertical and meridional structures of the moisture tendencies, but also  shifts in the intensity distribution of precipitation including extremes. Ablation experiments show that latent alignment improves both accuracy and the robustness across random seeds used in training. While some reduced skill remains in the boundary layer, the framework offers a data-driven alternative to manual feature engineering of climate invariant inputs. Beyond parameterizations used in hybrid ML-physics systems, the approach holds promise for other climate applications such as statistical downscaling. 
\end{abstract}

\section*{Plain Language Summary}
Predicting how the atmosphere will behave in a warmer world is one of the biggest challenges in climate science. While machine learning has shown promise in improving weather and climate models, most approaches struggle to work reliably outside the climate conditions they were trained on. 
In this study, we introduce CERA (Climate-invariant Encoding through Representation Alignment), a machine learning method designed to improve performance under climate change. CERA first learns a shared internal representation of the atmosphere using data from both present-day (control) and warmer ($+4\,^\circ$C) climates, without training on any outputs from the warmer climate. 
A second stage of the model then learns to predict key atmospheric processes using only outputs from the control climate. CERA makes more accurate predictions under warming than models that rely on raw inputs or manually engineered physical features.
CERA offers a flexible, data-driven alternative to previous approaches to climate-change generalization.
We apply CERA to moist physics processes in the atmosphere (including deep convection), but it holds potential for broader uses in climate science such as for estimating higher-resolution climate fields from coarse model output.

%
%

%


%
%
%
%

\section{Introduction}
Machine learning (ML) has emerged as a powerful tool for advancing weather and climate science \cite{Chantry2021OpportunitiesAI,Eyring2024PushingLearning, Schneider2022ESA-ECMWFPrediction,Nguyen2023Climax:Climate,Bracco2025MachineClimate}. It has been applied across a wide range of tasks, such as weather forcasting \cite{Weyn2021SubseasonalModels,Bi2022Pangu-weather:Forecast, Kurth2023Fourcastnet:Operators, Lam2023LearningForecasting, McNally2024DataObservations,Kochkov2024NeuralClimate,Price2025ProbabilisticLearning}, parameterization of subgrid processes \cite{OGorman2018UsingEvents,Gentine2018CouldDeadlock, Brenowitz2018PrognosticParameterization, Yuval2020Stable3295}, emulation of global climate models \cite{Cachay2024ProbabilisticDYffusion,Chapman2025CAMulator:Model,Watt-Meyer2025ACE2:Responses}, bias-correction of weather and climate models \cite{WattMeyer2021CorrectingSimulations,Bretherton2022CorrectingSimulations,Bora2023LearningOperators,Gregory2025AdvancingLearning,Chapman2025ImprovingCorrections} and statistical downscaling \cite{Rampal2022High-resolutionZealand, Hobeichi2023UsingDownscaling, Flora2025WoFSCast:Scales}. In particular, ML-based parameterizations have shown promise in representing the effects of complex subgrid processes such as moist convection and boundary-layer turbulence by training on high-resolution model data \cite{Brenowitz2018PrognosticParameterization,Yuval2020Stable3295,Han2023AnIntegration} or observations \cite{Zhao2019PhysicsconstrainedEvapotranspiration,McCandless2022MachineEstimates}.

Despite this promise, a persistent challenge for ML models in climate applications is their limited generalization to out-of-distribution conditions. This problem is especially acute under climate change, where shifts in temperature, moisture, and circulation patterns can push the system into regimes not well represented in the training data. For instance, ML parameterizations trained on present-day conditions often perform poorly when evaluated in warmer climates, leading to inaccurate predictions of precipitation and subgrid tendencies \cite{OGorman2018UsingEvents,Rasp2018DeepModels,Scher2019GeneralizationSystems}. This generalization gap limits the reliability and robustness of ML-based tools in climate change scenarios, precisely where improved predictions are most needed.

Several approaches have been proposed to improve performance in different climates. 
The most straightforward approach is to train ML models on high-resolution model output across multiple climates, so that the learned representation is implicitly exposed to climatic variability \cite{OGorman2018UsingEvents,Rasp2018DeepModels,Clark2022CorrectingSimulations,Sun2024DataWACCM,Bodnar2025ASystem}, although this implies a strong reliance on the quality of the high-resolution model output in different climates which is difficult to validate using observations. Also there is a need to ensure the training data covers a sufficiently wide range of climates for the application at hand. Another strategy is to incorporate physics-based constraints, such as conservation of energy and moisture, non-negativity of precipitation, or constraints based on known thermodynamic relationships which may improve generalization \cite{Brenowitz2020InterpretingConvection,Kashinath2021Physics-informedModelling,Yuval2021UsePrecision,Perezhogin2025GeneralizableModels}. 
In addition, \citeA{Beucler2024Climate-invariantLearning} proposed a climate-invariant feature engineering framework, in which input variables are transformed (e.g., from specific humidity and temperature to relative humidity and plume buoyancy) to reduce their distribution shift under climate forcing. While such physically motivated transformations have shown promise, they rely on hand-crafted features derived from expert knowledge. These transformations often require time-consuming trial-and-error tuning and may not yield optimal solutions, as they could sacrifice important information in pursuit of approximate climate invariance. Moreover, some approaches assume access to labeled outputs from future climates, which may not be available in practice, especially when using real-world observations.

In this study, we introduce a third path: learning climate-invariant structure directly from data, without relying on expert-designed features or supervision from warmer-climate labels. Our model consists of an autoencoder (AE) and a multilayer perceptron (MLP) trained jointly in an end-to-end fashion. The AE learns a latent representation of the input profiles, while the MLP predicts subgrid thermodynamic tendencies from this latent code. Inputs from both present-day and +4\,K warmer climates are used to train the AE, but the MLP is supervised only using control-climate outputs. To ensure that the latent space remains aligned across climates, we introduce a distributional regularization term based on the Earth Mover’s Distance (EMD) \cite{Rubner2000TheRetrieval}, which penalizes divergence between the latent encodings of control and +4\,K inputs.

The results show that our method achieves comparable or better performance than approaches based on hand-crafted physical inputs, without requiring any outputs from warmer climate conditions. We do still need inputs from the warmer climate, and in practice these could be obtained from climate-model simulations or potentially a pseudo-global warming approach applied to observations \cite{Schar1996SurrogateModels}. By not requiring outputs, and in particular the relationship between inputs and outputs, in the warmer climate, we are reducing dependence on high-resolution model output in a warmer climate. This highlights the potential of our method as a data-driven strategy for improving generalization in climate ML. 

We test our approach on the problem of parameterization of subgrid moist processes in the atmosphere using the neural-network parameterization of \citeA{Yuval2021UsePrecision}.
We consider a +4\,K warming as a stringent test of generalization, but we emphasize that a useful application in practice could be for a much smaller magnitude of warming. For example, one could train on recent decades and potentially use climate-invariance to improve generalization for a modest warming over the subsequent few decades or to improve robustness to extremes of internal variability that are not present in the training data.

We describe the data, architecture and baseline models in Section 2. We then describe the performance of CERA in Section 3 with an additional emphasis on precipitation which is important for climate impacts. Finally we discuss our results and their implications in Section 4.

\section{Methods}\label{sec3}
\subsection{Input–Output Specification and Experimental Setup}
We use the same high-resolution model output and coarse-graining procedure as in \citeA{Yuval2021UsePrecision} supplemented by a warmer simulation from \citeA{OGorman2021ResponseResolution} to test generalization across climates. Both simulations were performed using the System for Atmospheric Modeling (SAM, version 6.3) \cite{Khairoutdinov2003CloudSensitivities} in a quasi-global aquaplanet setup on an equatorial beta-plane with 48 vertical levels and a zonally symmetric SST distribution. A hypohydrostatic rescaling factor of 4 was used to enable convection-resolving simulations at 12 km horizontal resolution. The high-resolution simulation output was coarse-grained by a factor of 8 (to $96$ km) using spatial averaging. For a range of thermodynamic and moisture variables, subgrid fluxes were calculated by differencing coarse-grained and resolved fluxes, and coarse-grained tendencies were calculated by coarse-graining sources/sinks. The control simulation (0K) has baseline SSTs while the +4K warmer simulation has SSTs uniformly increased by 4K. These are the same SAM simulations as used in \citeA{Beucler2024Climate-invariantLearning} but in that study they were referred to as -4K and 0K. Each simulation is run for 600 days, with the final 500 days used for training and evaluation. The data from this period are randomly split into 95\% for training and 5\% for testing.

The inputs to the moist physics parameterization consist of vertical profiles of temperature $T$ and total non-precipitating water mixing ratio $q_T$, evaluated at the lowest 30 full model levels. This yields a total of $30 \times 2 = 60$ input features. The outputs include five vertically resolved quantities: the subgrid contribution to the flux of non-precipitating liquid/ice static energy $H_L$ due to vertical advection ($H_{L\_\mathrm{adv}}$) and the total tendency of $H_L$ from freezing and melting of precipitation ($H_{L\_\mathrm{phase}}$); the subgrid contribution to the flux of $q_T$ due to vertical advection ($q_{T\_\mathrm{adv}}$) and cloud ice sedimentation ($q_{T\_\mathrm{sed}}$), and the total tendency of $q_T$ due to microphysical conversions between $q_T$ and precipitating water ($q_{T\_\mathrm{mic}}$). These outputs are defined on different vertical grids: vertical advective fluxes are predicted at 29 ``half" model levels above the surface, ice sedimentation fluxes at 30 ``half" model levels, and total tendencies at 30 ``full" model levels. This results in a total of $29 \times 2 + 30 \times 3 = 148$ output features.

For full details of the simulation setup, coarse-graining methodology, and the energy and moisture variables used we refer the reader to \citeA{Yuval2021UsePrecision}. One deviation from that setup is that we omit radiative heating from the parameterization, as it is not clear a priori if it should have the same climate invariant inputs as the moist physics processes that are our main focus. Consequently, we exclude from the inputs the absolute meridional distance from the equator, which was previously used as a proxy for insolation, and we omit radiative heating from the outputs. Another difference is that our machine learning model uses input data that are normalized per column, whereas \citeA{Yuval2021UsePrecision} apply normalization per vertical level. Output variables are normalized per column, but for simplicity we do not apply the output reweighting for different types of physical variables used by \citeA{Yuval2021UsePrecision}.

However, we did include radiative heating in an alternative version of our analysis (results shown in Figure~\ref{fig.R2_2outputs}) which was designed to be directly comparable to \citeA{Beucler2024Climate-invariantLearning}.
The inputs in this case include the absolute meridional distance from the equator, $|y|$, which serves as a proxy for insolation. The five vertically resolved outputs listed earlier plus radiative heating are combined into the subgrid tendencies of $H_L$ and $q_T$ using equations S2 and S5 of \citeA{Yuval2021UsePrecision}. The model was retrained to directly predict these two combined tendencies.

We compare all our results with the climate invariant method as introduced in \citeA{Beucler2024Climate-invariantLearning}, in which the inputs are transformed to a plume buoyancy variable (B) and relative humidity (RH). These transformations were designed to improve the generalization of learned parameterizations across different climate states.

\subsection{Model Architecture, Training, and Evaluation}
\begin{figure}
	\centering
		\includegraphics[width=1\textwidth, trim=2 2 2 2,clip]{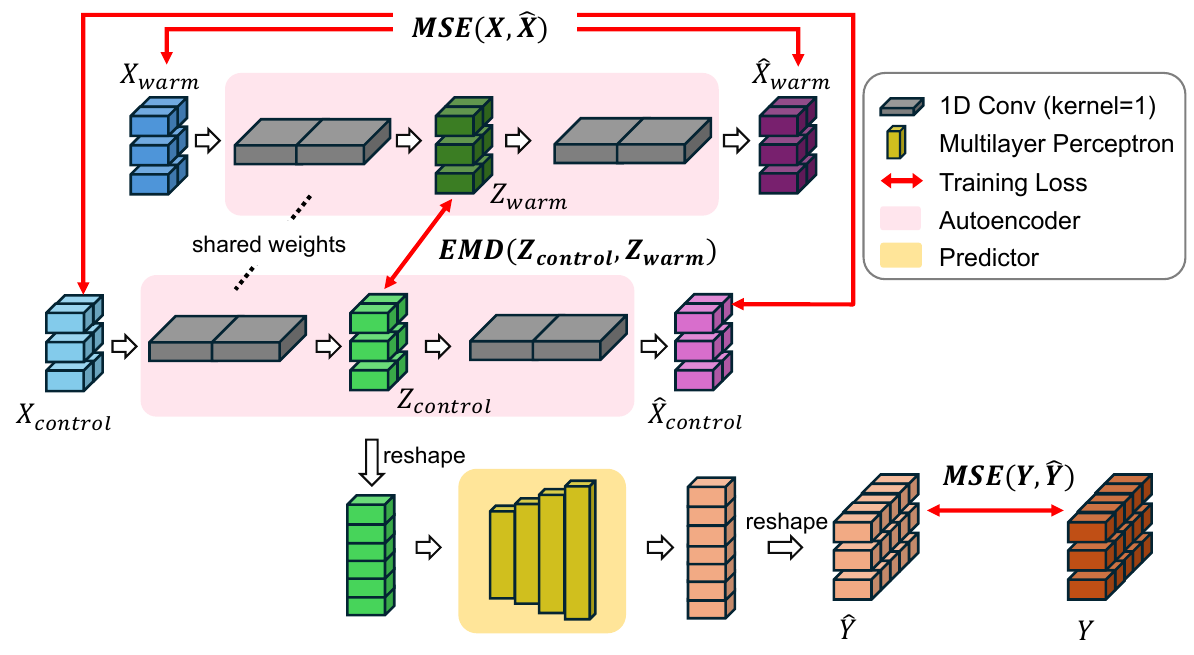}
	\caption{\textbf{Schematic of CERA.} The top half illustrates the autoencoder stage, where inputs from control ($X_\mathrm{cold}$) and warm ($X_\mathrm{warm}$) climates are encoded into latent representations ($Z_\mathrm{cold}$ and $Z_\mathrm{warm}$) using shared 1D convolution layers. Reconstruction losses $\mathrm{MSE}(X, \hat{X})$ are applied to both climates, while an Earth Mover’s Distance loss $\mathrm{EMD}(Z_\mathrm{cold}, Z_\mathrm{warm})$ encourages alignment between their latent spaces. The bottom half depicts the predictor stage, where the latent representation from the control climate is reshaped and passed through a predictor network to estimate target outputs $\hat{Y}$, with an $\mathrm{MSE}(Y, \hat{Y})$ loss applied for supervision. The framework enables learning a climate-invariant latent space, supporting generalization to warmer climate conditions.}
	\label{fig.model}
\end{figure}

We propose CERA (Climate-invariant Encoding through Representation Alignment), a self-supervised model that learns climate-invariant structure directly from raw inputs, without any feature engineering or labels from warmer climates.
As illustrated in Figure \ref{fig.model}, the model consists of two components: an autoencoder (AE) that processes multi-level input features (i.e., vertical profiles), and a multilayer perceptron (MLP) that predicts target outputs from the learned latent representations. 

The AE operates on a 60-dimensional input vector comprising vertical profiles of temperature and total non-precipitating water. The encoder uses a stack of one-dimensional convolutional layers. Each convolution has a kernel size of one and 64 channels, except for the final encoder layer, which outputs a three channel latent tensor $Z \in \mathbb{R}^{B \times 3 \times L}$, where $B$ is the batch size and $L$ is the number of vertical levels, which is 30 in our case. A kernel size of one means that each vertical level is transformed independently of its neighbors. This preserves vertical locality and can be viewed as applying the same dense transformation at each vertical level. 
Vertically non-local operations (e.g., larger kernels or attention across levels) could be explored in future work to capture vertical coupling. The decoder mirrors the encoder with transposed 1D convolutions, known as deconvolutions (see \citeA{Zeiler2010DeconvolutionalNetworks}) to reconstruct the original input from the full latent code $Z$, enabling self-supervised learning via reconstruction loss. 

To recognize that not all input information can be made exactly climate invariant while still reconstructing the inputs in different climates in the decoder, the latent space is partitioned. One channel of $Z$ at each vertical level is reserved for such non-aligned information for the AE and does not participate in either the distribution alignment or the downstream predictor. This partitioning is found to slightly improve the generalization for the outputs related to ice ($H_{L\_\mathrm{phase}}$ and $q_{T\_\mathrm{sed}}$). 

The MLP predictor takes only the aligned subset of $Z$ as input and maps it to a 148-dimensional output vector representing vertically resolved subgrid fluxes and tendencies. 
The MLP predictor is a five-layer fully connected neural network with 128 neurons per hidden layer and leaky ReLU activations. 

We adapted the model for the alternative version of our analysis (results shown in Figure~\ref{fig.R2_2outputs}) to ensure a direct comparison with \citeA{Beucler2024Climate-invariantLearning}, accounting for differences in input and output variables described in the previous section. Since the distance to the equator $|y|$ is assumed to affect only the radiative heating component, we process $|y|$ through a fully-connected two-layer MLP with 8 neurons in the hidden layer and 30 neurons in the output layer, and add its output as a residual correction exclusively to the energy tendency. This design choice minimizes the influence of $|y|$ on the rest of the network.

The AE is trained using input data from both the control and +4\,K warmer climate states, while the MLP is trained using only control-climate outputs as supervision. To enforce a shared latent structure across climate regimes, we introduced a loss term based on the Earth Mover’s Distance (EMD) \cite{Rubner2000TheRetrieval} between the latent distributions of control and +4K climate inputs. In one dimension, the EMD simplifies to the distance between sorted samples, allowing for an efficient closed-form computation \cite{Levina2001TheStatistics}. In our implementation, the EMD is computed separately for each latent channel at each vertical level and then averaged across all dimensions. This alignment mechanism forces the encoder to extract features that are common across both climates, effectively learning a climate-invariant representation space.

The model is trained end-to-end using a combined loss that includes reconstruction loss for the autoencoder, EMD loss to encourage latent alignment between climates, and prediction (pred) loss between predicted and true subgrid fluxes and tendencies in the control climate.
The total training loss is defined as:
\begin{equation}\label{eq_emd}
\mathcal{L}_{\text{total}} = (1 - \lambda_{\text{pred}} - \lambda_{\text{EMD}}) \cdot \mathcal{L}_{\text{reconstruction}} + \lambda_{\text{pred}} \cdot \mathcal{L}_{\text{pred}} + \lambda_{\text{EMD}} \cdot \text{EMD}(Z_0, Z_{+4K}),
\end{equation}
where:
\begin{itemize}
    \item $\mathcal{L}_{\text{reconstruction}}$ is the mean squared error between inputs and their reconstructions for both control and +4\,K climates,
    \item $\text{EMD}(Z_0, Z_{+4K})$ is the Earth Mover’s Distance between latent spaces from the two climate states,
    \item $\mathcal{L}_{\text{pred}}$ is the supervised loss (mean squared error) between predicted and ground-truth subgrid fluxes and tendencies, using control-climate labels, and
    \item $\lambda_{\text{EMD}}, \lambda_{\text{pred}}$ are tunable weights balancing the three objectives.
\end{itemize}

Further training details are provided in the Supplementary Materials. 

We compare CERA against three alternative models: (i) a baseline multilayer perceptron (MLP) trained on raw input profiles from the control climate (Baseline) \cite{Yuval2021UsePrecision}; (ii) a physically informed MLP trained on hand-crafted, climate-invariant features of relative humidity and plume buoyancy, following \citeA{Beucler2024Climate-invariantLearning} (RH+B baseline); and (iii) an ablation variant of CERA trained without the EMD alignment loss (CERA-noAlign). To assess robustness, all models were trained using five independent random seeds \cite{Haynes2023CreatingApplications}.

\section{Results}\label{sec2}

\subsection{Autoencoding and Latent Alignment Enable Climate-Invariant Representations}

\begin{figure}
	\centering
		\includegraphics[width=1\textwidth, trim=2 2 2 2,clip]{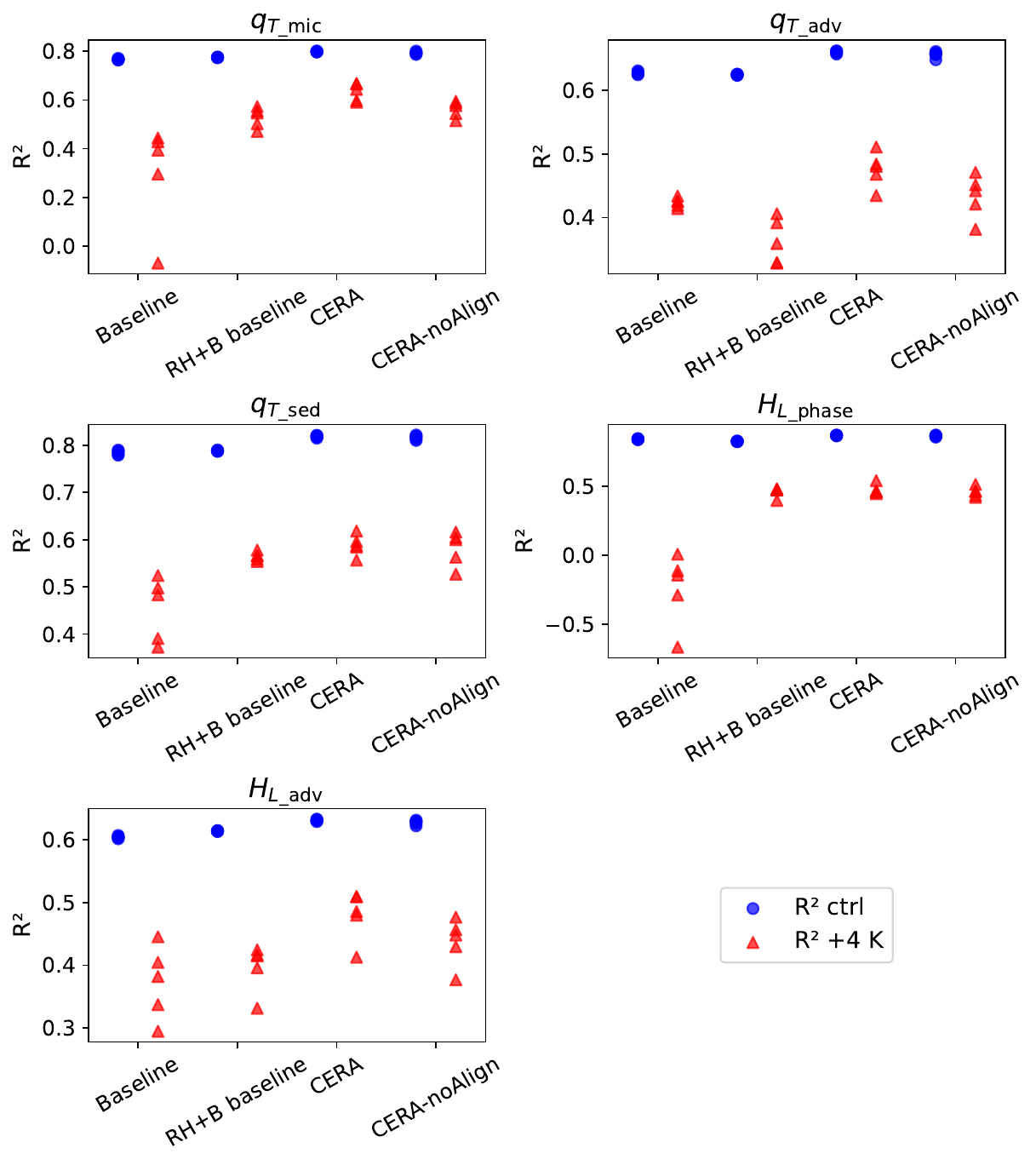}
	\caption{\textbf{R\textsuperscript{2} Comparison Across Models for Outputs in Control and $+4\,\mathrm{K}$ Climates.} Blue dots indicate the coefficient of determination ($R^2$) for each method (Baseline, RH+B baseline, CERA, and CERA-noAlign) in the control climate, evaluated across five subgrid variables. Red triangles show the corresponding $R^2$ scores under the $+4\,\mathrm{K}$ warming scenario. Results are displayed over five individual random seeds used in training per method. CERA achieves the highest overall performance, especially under warming, while RH+B exhibits greater robustness. CERA-noAlign improves upon the raw baseline but not to the same extent as CERA and with increased variability across random seeds, underscoring the importance of latent alignment. Note the different vertical axes for the different panels.}
	\label{fig.R2_5outputs}
\end{figure}

We evaluate the predictive performance of four models on five key subgrid tendency outputs (Figure \ref{fig.R2_5outputs}). Averaging $R^2$ over all output variables, CERA achieves the highest overall accuracy, with a mean $R^2$ of 0.75 in the control climate and 0.53 in the $+4\,\mathrm{K}$ climate. This indicates both strong in-distribution performance and substantial generalization capacity under warming.

In comparison, the Baseline model performs similarly to the RH+B baseline in the control climate (both have $R^2 = 0.72$) and slightly worse than CERA ($R^2=0.75$). That CERA outperforms the Baseline model in the control climate suggests that the restriction to using climate invariant inputs actually helps generalization beyond the training data even for test data in the control climate. The performance of the Baseline model deteriorates sharply under warming, with the mean $R^2$ falling to 0.26, highlighting a lack of ability to extrapolate beyond its training distribution as might be expected given distribution shifts in the input variables. The RH+B baseline maintains its performance to a much greater extent in the warm climate ($R^2 = 0.46$), outperforming the Baseline model but performing slightly worse than CERA ($R^2=0.53$).

These results underscore the effectiveness of the proposed latent-alignment strategy in enabling generalization across climates. Ablation experiments further emphasize its importance. The autoencoder-only variant (CERA-noAlign) improves upon the raw Baseline across all random seeds, indicating that learning a latent representation of the inputs helps the model extract more informative features for downstream prediction, even without the alignment. However, the addition of the EMD loss proves critical for achieving consistently high performance.

In addition to improving mean accuracy, CERA also enhances robustness across random seeds relative to the raw Baseline. The variance in $R^2$ among trained models is notably reduced for CERA compared to the Baseline and CERA-noAlign in the $+4\,\mathrm{K}$ climate. However, the RH+B baseline demonstrates the highest robustness overall. These results suggest that CERA may converge to different climate-invariant solutions across random seeds, leading to variability in performance. In contrast, the RH+B baseline represents a deterministic, hand-crafted solution that yields more consistent results across trained models. 
There may exist alternative hand-crafted input formulations capable of achieving higher accuracy. Identifying such configurations, however, would require extensive manual tuning and domain expertise, suggesting that self-supervised learning approaches like CERA may offer a more scalable alternative.

\begin{figure}
	\centering
		\includegraphics[width=1\textwidth, trim=2 2 2 2,clip]{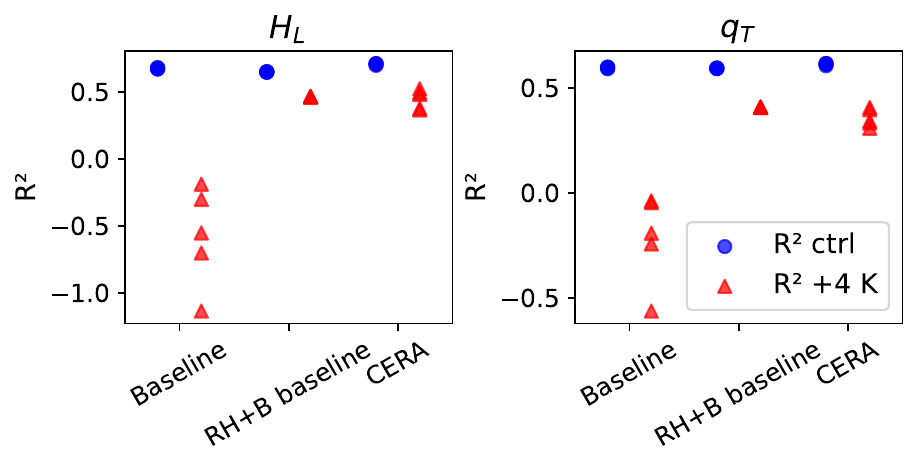}
	\caption{Same as Figure~\ref{fig.R2_5outputs}, but making a direct comparison to \citeA{Beucler2024Climate-invariantLearning} by retraining to include radiative heating in the parameterization and assessing skill for the tendencies of non-precipitating liquid/ice static energy $H_L$ and total non-precipitating water $q_T$.}
	\label{fig.R2_2outputs}
\end{figure}

To further assess performance, we make a direct comparison to \citeA{Beucler2024Climate-invariantLearning} by retraining to include radiative heating and assessing skill for the total subgrid liquid/ice static energy ($H_L$) and moistening ($q_T$) tendencies.
As shown in Figure~\ref{fig.R2_2outputs}, CERA achieves the highest $R^2$ scores in the control climate, outperforming both the raw-input Baseline and the RH+B baseline. 
In the $+4\,\mathrm{K}$ climate, CERA and RH+B produce comparable performance, both substantially outperforming the Baseline model. These results show that CERA is robust to changes in the outputs, and that again CERA not only generalizes effectively to a warmer climate, but it maintains high predictive skill in the training climate. 

\subsection{Evaluation of Performance for Precipitation}

Given the importance of precipitation for climate-change impacts, we next evaluate how well the ML models capture precipitation-related processes. In particular, we analyze the vertical profile of prediction skill for $q_{T\_\mathrm{mic}}$, the tendency of $q_T$ due to microphysical conversion to precipitation, and the resulting offline surface precipitation distributions. Details for computing the instantaneous surface precipitation rate can be found in the Supplementary Materials. All results were averaged across five random seeds per model.

\begin{figure}
	\centering
		\includegraphics[width=1\textwidth, trim=2 2 2 2,clip]{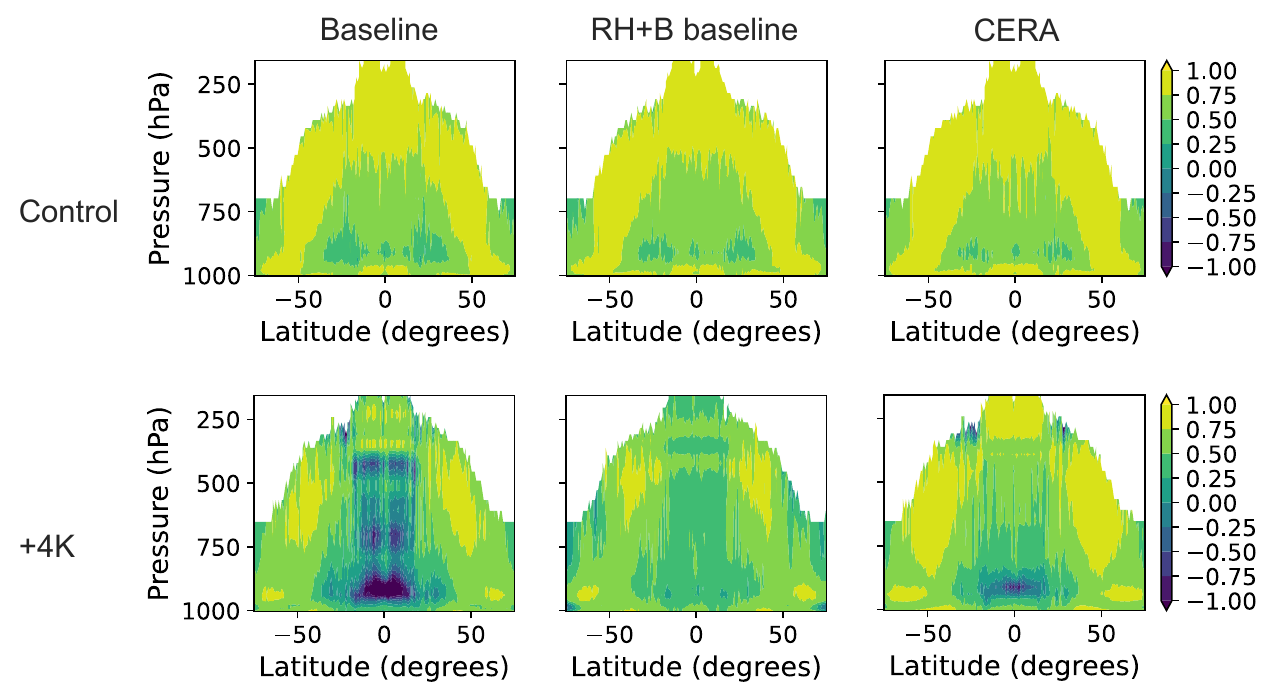}
	\caption{\textbf{$R^2$ for $q_{T\_\mathrm{mic}}$ versus latitude and pressure across methods and climates.} Shown are the coefficient of determination ($R^2$) values for predicted $q_{T\_\mathrm{mic}}$ (moisture tendency due to microphysical conversion to precipitating water) in the control (top) and $+4\,\mathrm{K}$ (bottom) climates, averaged over five random seeds. Each row compares three models: Baseline, RH+B baseline, and CERA. 
    Regions with variance less than 1\% of the mean variance across levels and latitudes were masked to remove near-constant areas. In the control climate, all models exhibit moderate-to-high skill. Under warming, performance differences become more pronounced. The Baseline model degrades sharply in the tropics, the loss in skill is much less in RH+B, while CERA retains high skill aloft though with some loss of skill in the boundary layer.}
	\label{fig.r2_qout}
\end{figure}

Figure~\ref{fig.r2_qout} shows $R^2$ for $q_{T\_\mathrm{mic}}$ versus pressure and latitude in the control and $+4\,\mathrm{K}$ climates. In the control climate, all three models have a similar skill structure with moderate-to-high skill in the mid-to-upper troposphere. In the $+4\,\mathrm{K}$ climate, performance disparities between methods becomes pronounced. The Baseline model suffers a substantial drop in predictive skill, with negative $R^2$ values in the tropics where it fails to extrapolate beyond its training climate. The RH+B baseline does not degrade to the same extent, but its skill is still substantially lower than in the control climate. In contrast, CERA maintains strong performance in the mid and upper troposphere but shows reduced skill in the boundary layer, particularly in the tropics. 

\begin{figure}
	\centering
		\includegraphics[width=1\textwidth, trim=2 2 2 2,clip]{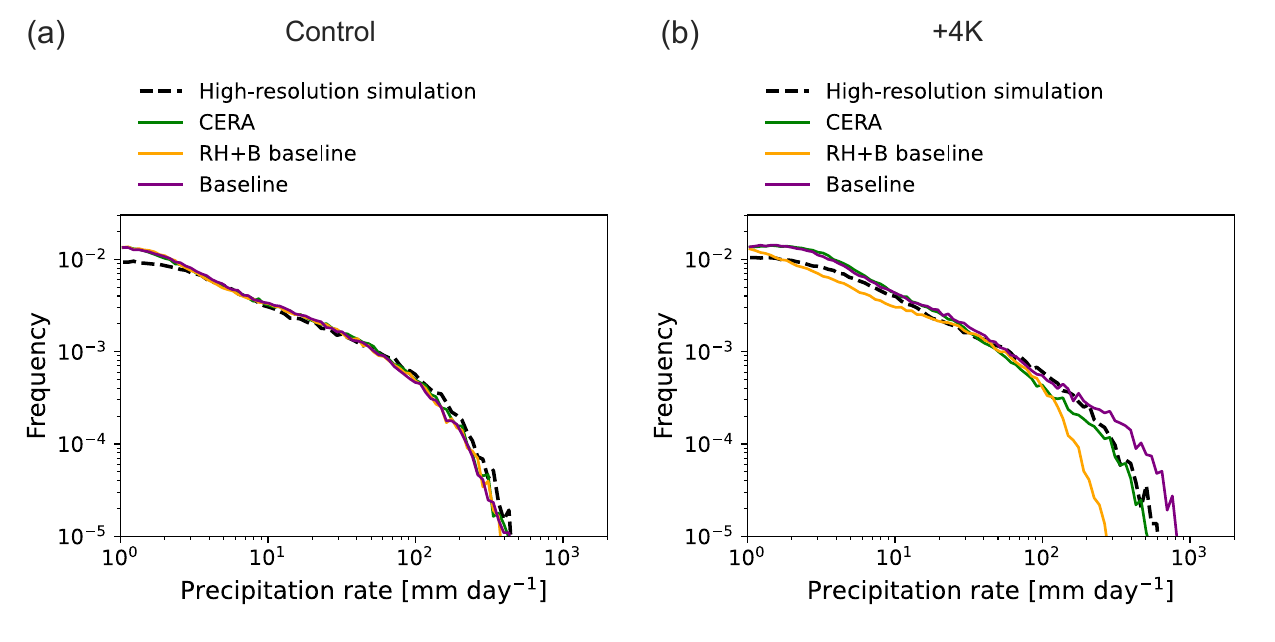}
	\caption{\textbf{Frequency distributions of instantaneous precipitation rates in the control and $+4\,\mathrm{K}$ climates.} Panel (a) shows distributions in the control climate, while panel (b) shows distributions under $+4\,\mathrm{K}$ warming.  Solid lines show model predictions averaged across five random seeds: purple for the Baseline, orange for RH+B, and green for CERA. The high-resolution reference simulation is shown in dashed black. Precipitation rates are derived from offline-integrated $q_{T\_\mathrm{mic}}$ and binned using 100 logarithmically spaced intervals starting at $1\,\mathrm{mm\,day^{-1}}$. All models are compared against the high-resolution simulation reference (ground truth). In the control climate, models generally reproduce the overall distribution shape but tend to overestimate drizzle ($1$--$2\,\mathrm{mm\,day^{-1}}$) and slightly underestimate extremes. Under warming, model performances diverge more substantially. The Baseline model overestimates across a broad range, RH+B underestimates throughout, and CERA closely tracks the reference in the tail but underpredicts the $40$--$300\,\mathrm{mm\,day^{-1}}$ range.
    }
	\label{fig.precip_dist}
\end{figure}

Figure~\ref{fig.precip_dist} shows the normalized frequency distributions of instantaneous precipitation rates under control and $+4\,\mathrm{K}$ warming scenarios. Precipitation rates are binned using 100 logarithmically spaced intervals starting at $1\,\mathrm{mm\,day^{-1}}$. Frequencies are normalized by the number of samples exceeding this threshold, such that the sum across all bins equals one, yielding a unitless vertical axis. These distributions provide a sensitive diagnostic of each model’s ability to reproduce the full range of precipitation intensities, from drizzle to extremes.

In the control climate (Figure~\ref{fig.precip_dist}a), all models exhibit similar performance and closely follow the reference distribution from the high-resolution simulation across most of the range. However, two systematic discrepancies are apparent. First, all models overestimate the frequency of light precipitation events in the $1$--$2\,\mathrm{mm\,day^{-1}}$ range. Second, all models slightly underestimate the frequency of extreme events above $\sim100\,\mathrm{mm\,day^{-1}}$. Despite these deviations, the overall agreement with the reference remains strong.

Under $+4\,\mathrm{K}$ warming (Figure~\ref{fig.precip_dist}b), more pronounced differences emerge. The Baseline model exhibits substantial overestimation across a wide range of intensities. Its bias for weak intensities becomes more pronounced, with an overestimation of precipitation below $4\,\mathrm{mm\,day^{-1}}$, and it substantially overestimates the frequency of extreme events above $200\,\mathrm{mm\,day^{-1}}$. In contrast, the RH+B baseline systematically underestimates the frequency of precipitation across almost all intensities. 
CERA more accurately reproduces the high-resolution reference, particularly at the extreme tail (above $300\,\mathrm{mm\,day^{-1}}$), where it tracks the enhanced frequency of intense events associated with warming. However, it tends to underestimate the frequency of precipitation in the range of $40$--$300\,\mathrm{mm\,day^{-1}}$. 

\begin{figure}
	\centering
		\includegraphics[width=1\textwidth, trim=2 2 2 2,clip]{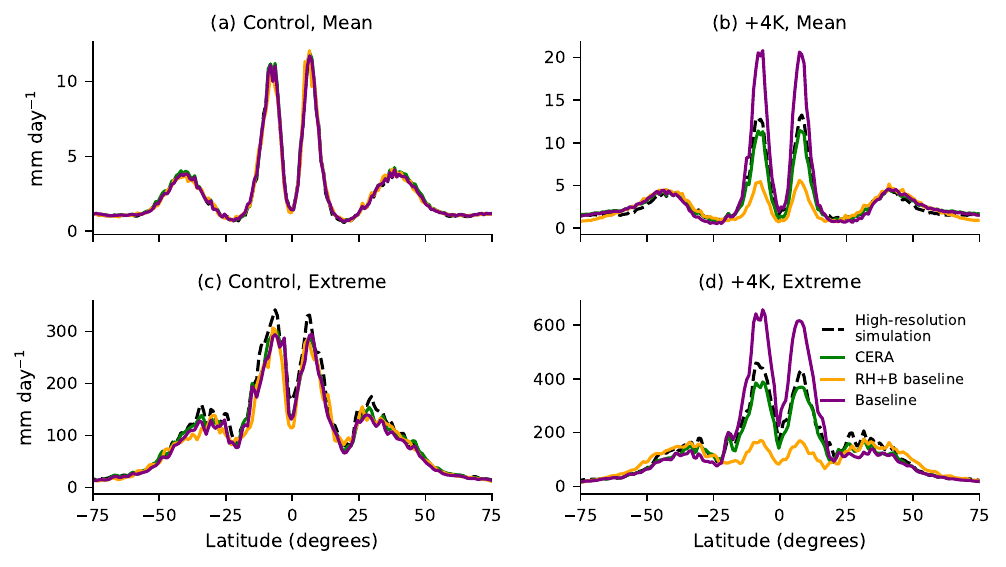}
	\caption{\textbf{Mean and extreme precipitation versus latitude in control and $+4\,\mathrm{K}$ climates.} Panels (a) and (b) show the mean precipitation for the control and $+4\,\mathrm{K}$ climates, respectively, while panels (c) and (d) display the corresponding 99.9th percentile (extreme) precipitation. All values are derived from offline-integrated $q_{T\_\mathrm{mic}}$. Solid lines represent model predictions averaged over five random seeds (purple for Baseline, orange for RH+B, green for CERA), and the dashed black line shows the high-resolution reference simulation. Extreme precipitation curves have been lightly smoothed using a 1-2-1 filter for readability.}
	\label{fig.zonal_precip}
\end{figure}

Further insight into model performance is provided by considering the precipitation distributions versus latitude shown in Figure~\ref{fig.zonal_precip}. In the control climate (Figure~\ref{fig.zonal_precip}a,c), all models reproduce the general meridional structures of the mean and extreme (99.9th percentile) precipitation. 
There are minor deviations in the extremes, with all models slightly underestimating the peak intensity.

In the $+4\,\mathrm{K}$ climate (Figure~\ref{fig.zonal_precip}b,d), model differences become pronounced in the tropics. CERA successfully captures both the intensification and the meridional structure of tropical precipitation, even if it exhibits a slight underestimation of both mean and extreme precipitation near the peak. In contrast, the Baseline model substantially overestimates the magnitude of tropical precipitation, particularly for extreme precipitation, while the RH+B baseline systematically underestimates both mean and extreme values across the tropics.

Together, these results demonstrate that CERA generalizes robustly across climates, effectively capturing the vertical and zonal structure of moist processes, although its performance degrades in the tropical boundary layer. The Baseline model shows a sharp decline in skill for +4K, particularly in the deep tropics, while the RH+B baseline has much less of a decline in skill for $q_{T\_\mathrm{mic}}$ but consistently underestimates precipitation throughout the tropics.

\section{Discussion}\label{sec3}
CERA demonstrates that combining autoencoding with latent space alignment offers a powerful and flexible approach for learning climate-invariant representations of subgrid processes. Unlike models that rely on hand-crafted input features, CERA learns directly from raw inputs, eliminating the need for manual feature engineering. This approach reduces development effort and avoids restrictive assumptions that could limit model skill or generalizability. 
Notably, CERA does not degrade performance in the control climate and in some cases even improves it, suggesting that latent alignment can enhance generalization without sacrificing in-distribution accuracy. A key future direction is to implement and evaluate CERA in online simulations. While this study focuses on moist-physics parameterization, the framework is broadly applicable and could be extended to other parameterized processes, or even to applications beyond parameterization. For example, CERA’s climate-invariant properties may benefit statistical downscaling tasks that demand robust extrapolation across climate regimes.

Nevertheless, the framework has limitations. While the latent alignment strategy enhances robustness, not all physical processes can be cast into climate-invariant forms. This imposes inherent limits on the generalizability of this approach. Future work could address these challenges by incorporating additional physical constraints or adapting the framework to account for climate-dependent processes. 

Additionally, the autoencoder in our framework employs one-dimensional convolutional layers with a kernel size of one, resulting in vertically local transformations that act independently at each level. While this design preserves interpretability, it precludes modeling vertical interactions. Future work could explore vertically non-local architectures to capture cross-level dependencies. Moreover, further analysis of the learned latent representations, such as applying symbolic regression, may also help uncover interpretable, low-dimensional expressions of key physical relationships.

\section*{Open Research Section}
To support transparency and reproducibility, the data and code are being prepared for public release and will be uploaded to GitHub and Zenodo.

\acknowledgments
Shuchang Liu was supported by SNSF Postdoc.Mobility Grant P500PN\_222180 and the MIT Climate Grand Challenge on Weather and Climate Extremes. Support was also provided by Schmidt Sciences, LLC. We thank BingXin Ke for fruitful discussions. We also thank Janni Yuval, Tom Beucler and the M2LInES team.

\newpage
\title{Supporting information of ``CERA: A Framework for Improved Generalization of Machine Learning Models to Changed Climates"}

%
%




\authors{Shuchang Liu\affil{1}, Paul A. O'Gorman\affil{1}}

\affiliation{1}{Department of Earth, Atmospheric, and Planetary Sciences, Massachusetts Institute of Technology}




\correspondingauthor{Shuchang Liu}{shuchang.liu@hotmail.com}

\newpage

\noindent\textbf{Contents of this file}
\begin{enumerate}
\item Training details
\item Formula for instantaneous surface precipitation rate
\end{enumerate}

\section{Training details} \label{training_details}
In each iteration, the autoencoder processes 8192 input samples, with 4096 from the control climate and 4096 from the $+4\,\mathrm{K}$ climate. The reconstruction loss is computed on the full batch, and the latent alignment loss is evaluated between the control and $+4\,\mathrm{K}$ samples. In the mean time, the downstream prediction loss is calculated using only labeled control-climate data. The final loss is a weighted sum of reconstruction loss, latent alignment loss and prediction loss to enable joint updates of the autoencoder and predictor.

The model is optimized using the AdamW optimizer with learning rates of $3 \times 10^{-3}$ and and a weight decay of $10^{-3}$ for both the autoencoder and the MLP predictor. A learning rate scheduler with exponential decay and 4000 linear warm up steps is applied to both learning rates. Training is run for 30 epochs.

After initial tests on generalization performance, hyperparameters were selected through a sweep aimed at balancing latent alignment and predictive accuracy. The weight on the latent alignment loss, $\lambda_{\text{EMD}}$, was chosen as the largest value that did not lead to latent space collapse, which we define as the average standard deviation of the latent representations falling below 0.1. Values of $\lambda_{\text{EMD}}$ ranging from $10^{-1}$ to $10^{-5}$ were tested, and $\lambda_{\text{EMD}} = 10^{-4}$ was selected as the largest stable value. 
The weight on the supervised prediction loss, $\lambda_{\text{pred}}$, was progressively decreased from 1 to 0.001. Since $\mathcal{L}_{\text{total}}$ depends on this weighting, we effectively selected $\lambda_{\text{pred}}$ by minimizing the unweighted combination of the underlying loss terms. For $\lambda_{\text{pred}}$ values between 1 and 0.01, $\mathcal{L}_{\text{pred}}$ varies by less than 5\% at the end of training, and $\mathcal{L}_{\text{reconstruction}}$ remains small ($< 10^{-4}$).
Interestingly, $\text{EMD}(Z_0, Z_{+4K})$ decreases when $\mathcal{L}_{\text{reconstruction}}$ becomes smaller, presumably because better reconstructions more strongly constrain the latent space. Consequently, $\text{EMD}(Z_0, Z_{+4K})$ reaches its minimum when $\lambda_{\text{pred}}$ is set to 0.01.
Further decreasing $\lambda_{\text{pred}}$ below 0.01 led to an increase in $\mathcal{L}_{\text{pred}}$. Therefore, the best-performing configuration used $\lambda_{\text{pred}} = 0.01$. This combination provided a stable latent alignment without compromising predictive performance on the labeled control-climate data.

For the alternative version of our analysis (results shown in Fig.~3) and to enable direct comparison with \citeA{Beucler2024Climate-invariantLearning}, we fine-tuned the $\lambda_{\text{EMD}}$ parameter to account for the modified input/output configuration.  We found that we could use a larger $\lambda_{\text{EMD}}$ to strengthen alignment, stopping before latent space collapse (defined as $\text{std} < 0.1$). We selected $\lambda_{\text{EMD}} = 10^{-3}$ as the largest value that preserved sufficient latent variability.

For the baseline models, they are the same as CERA without the autoencoder. We used the same learning rate ($3 \times 10^{-3}$ and and a weight decay of $10^{-3}$) for training the baseline models.

\section{Formula for instantaneous surface precipitation rate}

The instantaneous surface precipitation rate for both the ML models and the high-resolution simulation is computed by vertically integrating the microphysical tendency of total condensate, $q_{T\_\mathrm{mic}}$, with density weighting:
\begin{equation}
P_{\text{tot}}(z = 0) = -\int_0^{\infty} \rho_0 q_{T\_\mathrm{mic}}dz.
\end{equation}
For simplicity, we exclude the surface ice sedimentation flux, which is typically small. This formulation is similar to Equation S6 of \citeA{Yuval2021UsePrecision}, but we note that their equation omits a negative sign.

%
%


%
%
%
%
%
\bibliography{references.bib}

\end{document}